\documentclass{article}
\usepackage{spconf,amsmath,graphicx,hyperref,url,tablefootnote,subcaption}
\usepackage{booktabs,multirow,makecell}
\usepackage{xcolor}

\title{Augmenting Open-Vocabulary Dysarthric Speech Assessment with Human Perceptual Supervision}
\name{Kaimeng Jia*$^1$, Minzhu Tu*$^2$, Zengrui Jin$^1$, Siyin Wang$^1$, Chao Zhang$^1$ \thanks{* Equal contribution was made between the first two authors.}}
\address{
$^1$ Tsinghua University, $^2$ Beijing University of Posts and Telecommunications\\
\texttt{\small{jkm23@mails.tsinghua.edu.cn}}\small{,}
\texttt{\small{Epiphany\_1104@bupt.edu.cn}}\small{,} \\
\texttt{\small{wangsiyi23@mails.tsinghua.edu.cn}}\small{,}
\texttt{\small{\{zrjin,cz277\}@tsinghua.edu.cn}}
}
\begin{document}
%
\maketitle
\begin{abstract}
Dysarthria is a speech disorder characterized by impaired intelligibility and reduced communicative effectiveness. 
Automatic dysarthria assessment provides a scalable, cost-effective approach for supporting the diagnosis and treatment of neurological conditions such as Parkinson's disease, Alzheimer's disease, and stroke.
This study investigates leveraging human perceptual annotations from speech synthesis assessment as reliable out-of-domain knowledge for dysarthric speech assessment. 
Experimental results suggest that such supervision can yield consistent and substantial performance improvements in self-supervised learning  pre-trained models.
These findings suggest that perceptual ratings aligned with human judgments from speech synthesis evaluations represent valuable resources for dysarthric speech modeling, enabling effective cross-domain knowledge transfer.
\end{abstract}
\section{Introduction}
\label{sec:intro}

Dysarthria is a neuro-motor speech disorder resulting from neurological injuries or diseases, such as cerebral palsy, amyotrophic lateral sclerosis, Parkinson's disease, or stroke \cite{rudzicz2011tasl,rudzicz2009icassp}. 
These disruptions manifest in several perceptually salient characteristics, with reduced intelligibility and degraded naturalness being among the most prominent and impactful on daily communication.

The assessment of dysarthric speech, particularly in terms of intelligibility and naturalness, is of paramount importance due to its critical clinical and technological implications. This importance is reflected in several key areas of research: 
a) a sensitive biomarker for early detection and tracking of neurological disease progression \cite{10095664}, 
b) an objective measure to guide speech therapy strategies and monitor rehabilitation outcomes \cite{bhat2017icassp},
c) a key factor whose accurate assessment is crucial for improving the performance of automated systems, such as automatic speech recognition (ASR), and enabling technologies like disordered speech reconstruction \cite{jeon25_interspeech,10445949}.

Despite its critical importance, the clinical assessment of utterance-level intelligibility and naturalness relies heavily on subjective perceptual evaluations performed by certified speech pathologists \cite{rudzicz2012torgo,kim08c_interspeech}. 
The labor-intensive and time-consuming nature of subjective pathological speech evaluation significantly limit its accessibility and efficiency.
These limitations have motivated the development of automated, objective methods capable of providing scalable assessments. 
However, patients suffering from speech disorders are often compounded with co-occurring physical disabilities, lead to the difficulty in collecting large quantities of impaired speech required for model training and evaluation.
Moreover, existing approaches are restricted to assessments using a constrained lexicon and require matched control groups for pathological evaluation, resulting in an unnatural evaluation protocol. 
To the best of our knowledge, spontaneous utterances have not been employed in prior studies.


To this end, recent attention has been drawn towards achieving accurate automatic assessment of dysarthria despite the constraint of the data scarcity issue. 
Self-supervised learning (SSL) has emerged as a powerful paradigm in speech processing, offering robust and transferable representations from large-scale unlabeled corpora. 
Recent works such as wav2vec 2.0 \cite{baevski2020wav2vec}, HuBERT \cite{hsu2021hubert, yang2024k2ssl} have been successfully applied to both detection and severity assessment of dysarthria \cite{yeo2023icassp, javanmardi2023wav2vec}. 

A wide spectrum of techniques has also been explored in dysarthric speech processing, including but not limited to, adversarial domain adaptation for generalization on unseen speakers \cite{10889800}, data augmentation utilizing speech synthesis facilitated by Diffusion Probabilistic Models \cite{wang23qa_interspeech} or by reverse autoencoders transforming healthy speech into dysarthric speech for model training \cite{bhat22_interspeech}, and other pre-trained TTS systems \cite{hermann23_interspeech,leung24_interspeech,kim25w_interspeech}. 
However, existing generative approaches focus on either modeling the fine-grained spectro-temporal characteristics of dysarthric speech for speech recognition system construction, or reconstruction of impaired speech. 
None of the prior works provide supervision precisely aligns with human perception.

Recently, automated assessment of Text-to-Speech (TTS) synthesis quality has been extensively investigated \cite{huang22f_interspeech}, producing a large amount of human annotated utterances with either fine-grained perceptual labels with accurate timestamps, or utterance-level scores measuring naturalness and intelligibility following the Mean Opinion Score (MOS) scoring protocol. 
Leveraging such TTS-derived perceptual annotations is particularly valuable for dysarthric speech assessment, since these measures closely mirror how human listeners judge degraded speech and can therefore serve as reliable supervision signals for data augmentation.

The contribution of this work is therefore threefold:
\begin{itemize}
	\item An empirical study of utilizing pre-trained SSL features for automatic dysarthric speech assessment is conducted, focusing on utterances recorded with unconstrained lexicon.
    \item Experiments suggested that synthesized speech with human annotated scores could effectively augment the performance of dysarthric speech assessment systems.
    \item Results suggest that failures in speech synthesis and manifestations of dysarthria share perceptual and acoustic commonalities, offering a novel perspective on how speech generation errors and articulatory disorders converge.
\end{itemize}

The rest of this paper is organized as follows. 
Section \ref{sec:task} introduces the automatic assessment task and describes the two supporting datasets: the SAP and QualiSpeech corpora.
Section \ref{sec:method} proposes method, including the model architecture and the two training paradigms: joint training and fine-tuning. 
Section \ref{sec:exp} presents the experiments and results on both intelligibility and naturalness prediction tasks. 
The last section concludes this work.

\section{Task Description}
\label{sec:task}

\subsection{The Speech Accessibility Project Challenge 2025}

\begin{figure}[ht]
  \centering
  \begin{subfigure}[b]{0.49\columnwidth}
    \centering
    \includegraphics[width=\columnwidth]{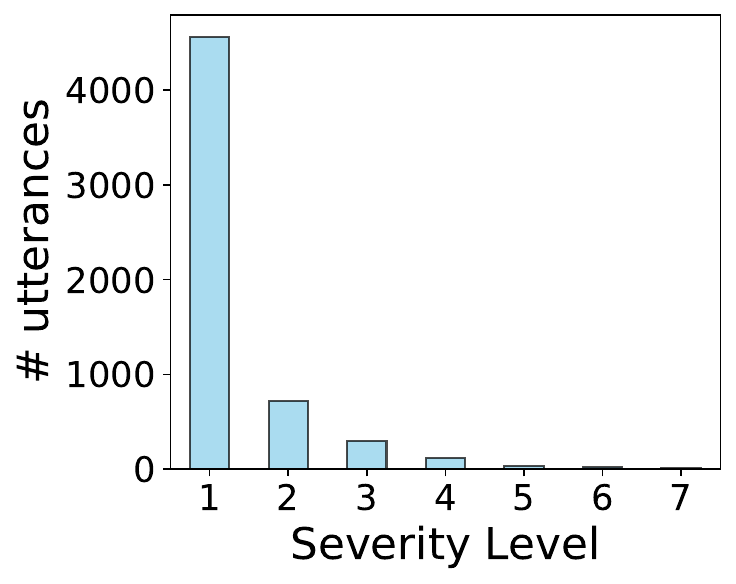}
    \caption{Intelligibility}
    \label{fig:naturalness_dist}
  \end{subfigure}
  \hfill
  \begin{subfigure}[b]{0.49\columnwidth}
    \centering
    \includegraphics[width=\columnwidth]{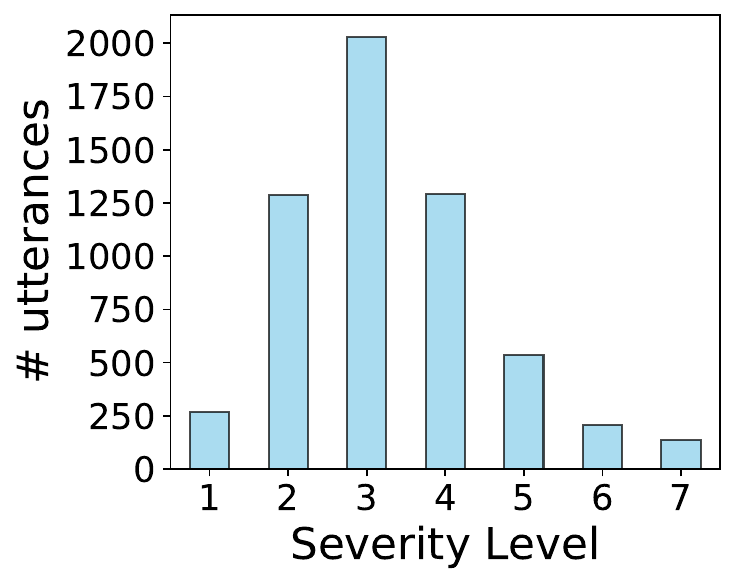}
    \caption{Naturalness}
    \label{fig:intelligibility_dist}
  \end{subfigure}
  \caption{Distribution of utterances across severity levels for (a) Intelligibility and (b) Naturalness in the Speech Accessibility Project (SAP) challenge.}
  \label{fig:distribution}
\end{figure}

The Speech Accessibility Project (SAP) \cite{zheng25_interspeech} challenge provides a large-scale, open-domain corpus of dysarthric speech.
The corpus comprises utterances recorded from over 500 speakers diagnosed with Parkinson's disease, Down syndrome, amyotrophic lateral sclerosis, Cerebral palsy, or stroke, encompassing more than 400 hours of speech and over 190,000 utterances.
The dataset was recorded through participants using their personal devices at home, capturing both read and spontaneously generated speech with unconstrained vocabulary to ensure naturalness and variability.
Fine-grained perceptual speech assessment across multiple clinically relevant dimensions are provided in the corpus, including but not limited to, harsh voice, inappropriate silences, pitch level, naturalness, variable rate, and intelligibility. 
Each dimension is annotated by certified speech-language pathologists using a 7-point scale, where a higher score indicates more significant severity is manifested in the corresponding dimension. 
Given the complexity among these perceptual dimensions and the primary objective of dysarthric speech assessment, the focus is specifically narrowed to two core utterance-level annotations: Intelligibility and Naturalness.
Distribution of utterances across severity levels can be seen from Figure \ref{fig:distribution}.

Due to the limited availability of the original SAP test set, the development set of the corpus was repurposed as the test set. 
For each dimension, the training set comprises utterances specifically annotated for that dimension; accordingly, for Naturalness, 5,040 training utterances were obtained, while for Intelligibility, 5,046 utterances were selected. 
Validation sets consisting of 500 utterances each were randomly sampled from the corresponding training subsets. 
The resulting test sets include 714 utterances for Naturalness and 716 utterances for Intelligibility.

\begin{figure*}[ht]
  \centering
  \includegraphics[width=\textwidth]{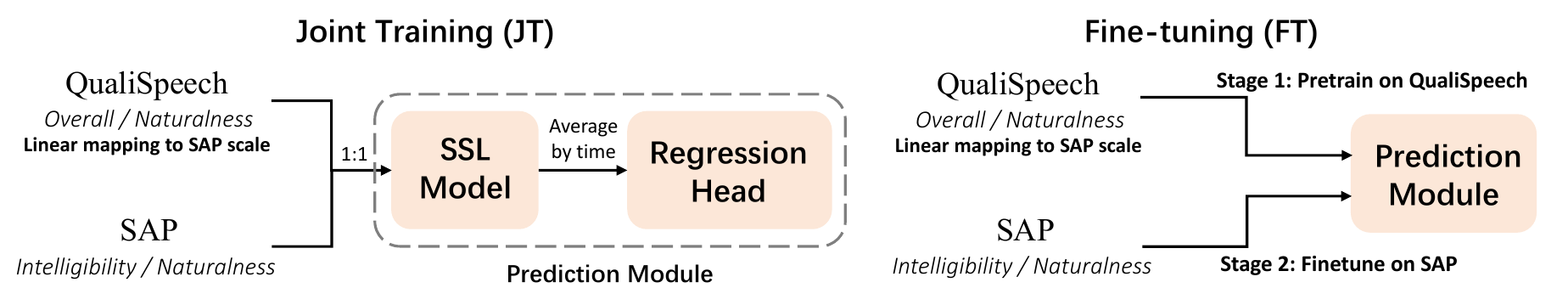}
  \caption{An illustration of the proposed joint training and fine-tuning paradigms, conducted and evaluated independently, integrating both QualiSpeech and Speech Accessibility Project (SAP) corpora for dysarthria severity prediction.}
  \label{fig:framework}
\end{figure*}

\subsection{QualiSpeech: A Descriptive Corpus for Speech Quality Assessment}
QualiSpeech \cite{wang-etal-2025-qualispeech} is an English corpus for low-level speech quality assessment with comprehensive annotations and detailed descriptive comments. 
It comprises three speech categories: synthetic speech (49\%), primarily from BVCC \cite{cooper21_ssw} and including various TTS models with standardized MOS scores; simulated real speech (27\%), sourced from NISQA \cite{mittag21_interspeech} featuring distortions that emulate actual transmission; and real speech (24\%), encompassing NISQA live recordings and in-the-wild samples from GigaSpeech \cite{chen21o_interspeech}. 
To ensure balanced speech quality conditions, 20\% of the synthetic speech samples are mixed with noise at signal-to-noise ratios ranging between 0 and 15 dB, while real speech samples are stratified into four quality groups based on UTMOS-predicted MOS scores \cite{saeki22c_interspeech}. 
Each utterance is evaluated across seven perceptual quality dimensions: noise, distortion, speed, continuity, listening effort, naturalness, and overall quality.
This protocol results in training, validation, and test splits of 10,558, 2,167, and 1,852 utterances, respectively, with balanced compositions of synthetic and real speech. 

Among the seven annotated dimensions, overall quality and naturalness are selected as they provide complementary yet comprehensive perspectives on perceptual evaluation. 
Overall quality captures the aggregated listener impression across multiple degradations, while naturalness reflects the degree to which speech resembles genuine human production. 
Focusing on these two dimensions ensures both clinical relevance for pathological speech assessment and consistency with established practices in general speech quality evaluation.

\begin{table}
  \centering
  \caption{Details of self-supervised learning (SSL) pre-trained encoders.}
  \label{tab:fairseq_models}
  \resizebox{\linewidth}{!}{
  \begin{tabular}{l c l}
    \toprule
    Model &  \# Param. & Dataset   \\
    \midrule
    wav2vec 2.0 Base\tablefootnote{\url{https://dl.fbaipublicfiles.com/fairseq/wav2vec/wav2vec_small.pt}} & 94M  & Librispeech \cite{panayotov2015librispeech}      \\
    wav2vec 2.0 Large*\tablefootnote{\url{https://dl.fbaipublicfiles.com/fairseq/wav2vec/wav2vec_vox_new.pt}} & 315M & Libri-Light \cite{kahn2020libri}  \\
    wav2vec 2.0 Large+\tablefootnote{\url{https://dl.fbaipublicfiles.com/fairseq/wav2vec/w2v_large_lv_fsh_swbd_cv.pt}} & 315M & \makecell[l]{Libri-Light \cite{kahn2020libri} + CommonVoice \cite{ardila-etal-2020-common} + \\ Switchboard \cite{godfrey1992switchboard} + Fisher \cite{cieri-etal-2004-fisher}} \\
    \midrule
    HuBERT Base\tablefootnote{\url{https://dl.fbaipublicfiles.com/hubert/hubert_base_ls960.pt}} & 95M  & Librispeech \cite{panayotov2015librispeech}  \\
    HuBERT Large\tablefootnote{\url{https://dl.fbaipublicfiles.com/hubert/hubert_large_ll60k.pt}} & 316M    & Libri-Light \cite{kahn2020libri} \\
    \bottomrule
  \end{tabular}
  }
\end{table}

\section{Method}
\label{sec:method}

\subsection{Model Architecture}


In order to leverage out-of-domain knowledge \cite{cooper2022generalization}, self-supervised learning (SSL) pre-trained encoders are adopted for feature extraction.
Frame-level features extracted from an SSL encoder are mean-pooled over time to obtain the utterance-level embedding, which are then fed into a regression head consisting of a two-layer feed-forward network with ReLU activation and dropout regularization applied between them, producing a single severity score. 

\subsection{Training Paradigms}



Two training paradigms, joint training (JT) and fine-tuning (FT), are evaluated separately to leverage perceptual annotations from QualiSpeech, as illustrated in Figure \ref{fig:framework}.
The JT paradigm employs a unified framework that simultaneously integrates the QualiSpeech and SAP corpora, balancing the two datasets by randomly sampling utterances from the QualiSpeech training set to match the size of the SAP training split, thus yielding a 1:1 ratio. 
In contrast, the FT paradigm involves pre-training the model on QualiSpeech, followed by fine-tuning on the SAP corpus.

Due to varying rating scales between QualiSpeech (MOS on a 1–5 scale) and SAP (severity ratings on a 1–7 scale), a linear transformation is applied to align scores across the datasets under the JT paradigm. 
Under this mapping, a QualiSpeech MOS of 5, representing the highest perceived quality, corresponds to an SAP score of 1, indicating utterance manifesting no dysarthric characteristics.

\subsection{Evaluation Metrics}
Model performance is evaluated using three standard metrics. 
Mean Squared Error (MSE), Linear Correlation Coefficient (LCC), and Spearman's Rank Correlation Coefficient (SRCC). 
MSE measures the deviation between predicted severity and ground-truth ratings, with lower values indicating smaller errors. 
LCC assesses the strength of linear association, while SRCC evaluates the consistency of ranking order. 
Higher LCC and SRCC values reflect stronger agreement with human subjective evaluations.

\begin{table*}[ht]
  \centering
  \renewcommand{\arraystretch}{0.9}
  \setlength{\tabcolsep}{3pt}

  \caption{Results of self-supervised learning (SSL) pre-trained encoders under joint training (JT) and fine-tuning (FT) on SAP and QualiSpeech. ``IDT'' denotes in-domain training on SAP only. For each SSL encoder, Mean Squared Error (MSE $\downarrow$), Linear Correlation Coefficient (LCC $\uparrow$), and Spearman's Rank Correlation Coefficient (SRCC $\uparrow$) are reported. ``Dimension'' lists the annotated attributes from the corresponding dataset. ``SAP'' stands for the Speech Accessibility Project dysarthric speech corpus.}
  \label{tab:qs_sap_results}

  \resizebox{\linewidth}{!}{
  \begin{tabular}{cc c *{5}{ccc}}
    \toprule
    \multicolumn{2}{c}{\textbf{Dimension}} &
    \multirow{2}{*}{\textbf{Method}} &
    \multicolumn{3}{c}{wav2vec 2.0 Base} &
    \multicolumn{3}{c}{wav2vec 2.0 Large*} &
    \multicolumn{3}{c}{wav2vec 2.0 Large+} &
    \multicolumn{3}{c}{HuBERT Base} &
    \multicolumn{3}{c}{HuBERT Large} \\
    \cmidrule(lr){4-6} \cmidrule(lr){7-9} \cmidrule(lr){10-12} \cmidrule(lr){13-15} \cmidrule(lr){16-18}
    SAP & QualiSpeech &
       & MSE & LCC & SRCC
       & MSE & LCC & SRCC
       & MSE & LCC & SRCC
       & MSE & LCC & SRCC
       & MSE & LCC & SRCC \\
    \midrule

    \multirow{5}{*}{Intelligibility} & -- & IDT
        & 0.348 & 0.628 & \textbf{0.482}
        & 0.421 & 0.523 & 0.368
        & 0.547 & 0.471 & 0.322
        & 0.461 & 0.540 & \textbf{0.406}
        & 0.475 & 0.485 & 0.404 \\
      & \multirow{2}{*}{Overall quality} & FT
        & \textbf{0.272} & \textbf{0.751} & 0.427
        & \textbf{0.351} & \textbf{0.612} & 0.393
        & \textbf{0.308} & 0.534 & 0.285
        & 0.487 & \textbf{0.594} & 0.368
        & 0.431 & 0.596 & 0.356 \\
      &  & JT
        & 0.408 & 0.590 & 0.446
        & 0.560 & 0.317 & 0.330
        & 0.627 & 0.225 & 0.302
        & 0.612 & 0.387 & 0.317
        & 0.526 & 0.479 & 0.363 \\
    \cmidrule(lr){2-18}
      & \multirow{2}{*}{Naturalness} & FT
        & 0.303 & 0.648 & 0.464
        & 0.387 & 0.572 & \textbf{0.401}
        & 0.495 & \textbf{0.566} & \textbf{0.443}
        & \textbf{0.379} & 0.451 & 0.295
        & \textbf{0.379} & \textbf{0.608} & \textbf{0.464} \\
      &  & JT
        & 0.433 & 0.602 & 0.475
        & 0.604 & 0.383 & 0.352
        & 0.528 & 0.445 & 0.348
        & 0.489 & 0.515 & 0.309
        & 0.451 & 0.551 & 0.391 \\

    \midrule

    \multirow{5}{*}{Naturalness} & -- & IDT
        & 1.127 & 0.581 & 0.591
        & 1.354 & 0.469 & 0.481
        & 1.119 & 0.574 & 0.607
        & 1.169 & 0.546 & 0.521
        & 0.941 & 0.570 & 0.503 \\
      & \multirow{2}{*}{Overall quality} & FT
        & 1.053 & \textbf{0.718} & \textbf{0.723}
        & 1.033 & 0.695 & 0.706
        & 1.000 & 0.686 & \textbf{0.696}
        & \textbf{0.847} & 0.644 & 0.587
        & \textbf{0.819} & \textbf{0.701} & \textbf{0.690} \\
      &  & JT
        & 0.909 & 0.691 & 0.680
        & 1.027 & 0.606 & 0.613
        & 1.106 & 0.589 & 0.617
        & 1.050 & 0.614 & 0.634
        & 0.930 & 0.661 & 0.668 \\
    \cmidrule(lr){2-18}
      & \multirow{2}{*}{Naturalness} & FT
        & \textbf{0.717} & 0.717 & 0.657
        & \textbf{0.800} & \textbf{0.709} & \textbf{0.713}
        & \textbf{0.880} & \textbf{0.692} & 0.690
        & 1.075 & \textbf{0.646} & \textbf{0.635}
        & 0.823 & 0.678 & 0.675 \\
      &  & JT
        & 1.022 & 0.637 & 0.643
        & 1.085 & 0.629 & 0.648
        & 1.022 & 0.637 & 0.642
        & 1.121 & 0.586 & 0.590
        & 1.036 & 0.638 & 0.635 \\
    \bottomrule
  \end{tabular}
  }
\end{table*}

\section{Experiments and Results}
\label{sec:exp}

In order to enhance the performance of automatic dysarthria assessment, joint training (JT) and fine-tuning (FT) paradigms are conducted and evaluated separately, to leverage human perceptual aligned supervision from QualiSpeech. 
All models evaluated are optimized using Adam optimizer with a learning rate of $1e-5$, weight decay of $0.01$. 
Details of the utilized self-supervised learning (SSL) pre-trained encoders are listed in Table \ref{tab:fairseq_models}.
All training paradigms share an identical experimental setup with parameters of SSL encoders tuneable.
Under the JT paradigm, 4,000 utterances are randomly sampled from the QualiSpeech training set with the original data distribution preserved, and then combined with the SAP training set for joint optimization.

\subsection{Performance on the Intelligibility Dimension}

Table \ref{tab:qs_sap_results} (Row 1–5) reports intelligibility prediction on SAP, with Overall quality and Naturalness from QualiSpeech used as auxiliary supervision under fine‑tuning (FT) and joint training (JT) across five SSL encoders.
Several trends can be observed: 
(i) wav2vec 2.0 Base consistently achieves superior performance under the in-domain training (IDT) paradigm (Row 1), yielding a 36.4\% relative MSE reduction compared to Large+, indicating better generalization capabilities of smaller encoders for this specific task;
(ii) Overall quality from QualiSpeech (FT) consistently enhances performance across all wav2vec encoders relative to IDT (Row 2 {\it vs.} 1), achieving the largest relative MSE improvement of 43.7\% for Large+ and a notable correlation increase of 19.6\% for Base;
(iii) Within intelligibility prediction tasks, FT consistently outperforms JT across encoders (Row 2 {\it vs.} 3, Row 4 {\it vs.} 5); wav2vec 2.0 Base emerges as the best-performing encoder within each comparison, Overall quality provides the most robust MSE and LCC improvements (Row 2 {\it vs.} 1), whereas Naturalness further contributes to SRCC improvements, particularly benefiting larger encoders (Row 4 {\it vs.} 1).

\subsection{Performance on Naturalness Dimension}
Table \ref{tab:qs_sap_results} (Row 6–10) reports naturalness prediction on SAP, utilizing Overall quality and Naturalness from QualiSpeech as auxiliary supervision under fine-tuning (FT) and joint training (JT) across five SSL encoders. Several trends can be observed:
(i) Overall quality from QualiSpeech (FT, Row 7 {\it vs.} 6) consistently enhances performance across all encoders, with wav2vec 2.0 Large* exhibiting the largest LCC (48.2\% relative) and SRCC (46.8\% relative) improvements, and HuBERT Base showing a relative MSE reduction of 27.6\%;
(ii) Overall quality under JT (Row 8 {\it vs.} 6) also outperforms IDT, albeit less effectively than FT, demonstrated by wav2vec 2.0 Base reducing MSE by 19.4\% and wav2vec 2.0 Large* by 24.2\%, along a 32.8\% SRCC improvement given by HuBERT Large;
(iii) Naturalness under FT (Row 9) consistently yields the lowest overall errors, with wav2vec 2.0 Base and Large* achieving a 36.4\% and 40.9\% relative MSE reduction, respectively.

\section{Conclusion}
This study demonstrated that leveraging perceptual annotations from the QualiSpeech corpus significantly enhances automatic dysarthria assessment. 
Both proposed augmentation methods, fine-tuning and joint training, improved prediction accuracy over the baseline, with fine-tuning consistently outperforming joint training. 
Larger models derived greater benefit from additional supervision, and naturalness predictions proved more reliable than intelligibility predictions. 
The effectiveness of synthetic speech augmentation underscores perceptual and acoustic commonalities between synthesis failures and dysarthric speech, highlighting promising directions for future clinical research.

\vfill\pagebreak
\bibliographystyle{IEEEbib}
\bibliography{refs}

\end{document}